\documentclass[aps,prd,12pt,nofootinbib,superscriptaddress]{revtex4}
\usepackage{graphicx}
\usepackage{amsmath}
\usepackage{amssymb}
\usepackage{mathrsfs}
\usepackage{verbatim}
\usepackage{dutchcal}
\usepackage{comment}
\usepackage{float}
\usepackage[normalem]{ulem}
\usepackage{xcolor}
\usepackage{slashed}
\newcommand{\bea}{\begin{eqnarray}}
\newcommand{\eea}{\end{eqnarray}}
\newcommand{\be}{\begin{equation}}
\newcommand{\ee}{\end{equation}}
\newcommand{\re}[1]{(\ref{#1})}

\newcommand{\stkout}[1]{\ifmmode\text{\sout{\ensuremath{#1}}}\else\sout{#1}\fi}

\begin{document}

\title{Fermions localized on $U(1)$ gauged Q-balls
}

\author{
Vladimir Dzhunushaliev
}
\email{v.dzhunushaliev@gmail.com}
\affiliation{
Department of Theoretical and Nuclear Physics,  Al-Farabi Kazakh National University, Almaty 050040, Kazakhstan
}
\affiliation{
Institute of Nuclear Physics, Almaty 050032, Kazakhstan
}
\affiliation{
Academician J.~Jeenbaev Institute of Physics of the NAS of the Kyrgyz Republic, 265 a, Chui Street, Bishkek 720071, Kyrgyzstan
}
\affiliation{
International Laboratory for Theoretical Cosmology, Tomsk State University of Control
Systems and Radioelectronics (TUSUR),
Tomsk 634050, Russia
}

\author{Vladimir Folomeev}
\email{vfolomeev@mail.ru}
\affiliation{
Institute of Nuclear Physics, Almaty 050032, Kazakhstan
}
\affiliation{
Academician J.~Jeenbaev Institute of Physics of the NAS of the Kyrgyz Republic,
265 a, Chui Street, Bishkek 720071, Kyrgyzstan
}
\affiliation{
International Laboratory for Theoretical Cosmology, Tomsk State University of Control
Systems and Radioelectronics (TUSUR),
Tomsk 634050, Russia
}

\author{Yakov Shnir}
\email{shnir@theor.jinr.ru}
\affiliation{BLTP, JINR, Dubna 141980, Moscow Region, Russia}
\affiliation{Instituto de F\'{i}sica de S\~{a}o Carlos; IFSC/USP;
Universidade de  S\~{a}o Paulo, USP\\
Caixa Postal 369, CEP 13560-970, S\~{a}o Carlos-SP, Brazil}


\begin{abstract}
The axially symmetric $U(1)$ gauged self-interacting Q-balls are shown to support normalizable fermionic bound state, minimally coupled to the electromagnetic field of the Q-ball. It is shown that the effects of the backreaction of the fermionic mode are very small.
We explore the domain of existence of
the solutions and address some of their physical properties. Among other properties, we observe that the eigenvalue of the Dirac operator remains positive, there is no zero mode in the spectrum.
\end{abstract}


\maketitle
\newpage
\section{Introduction}
The fermionic  modes dynamically localized on solitons in various spacetime dimensions
have been
the subject of much interest in past few decades. They were first
discussed in the pioneering paper \cite{zeromode} and later
revisited in Ref.~\cite{Jackiw:1975fn}. There has been a great deal of interest in
studying the localized fermionic states in various systems;
examples of such are fermion modes of the vortices \cite{deVega:1976rt,Jackiw:1981ee}, sphalerons
\cite{Boguta:1985ut}, superconducting cosmic strings \cite{Semenoff:1987ki,deSousaGerbert:1988kpn},
monopoles \cite{Rubakov:1982fp,Callan:1982au}, and
skyrmions \cite{Kahana:1984dx,Ripka:1985am}.

The typical assumption in most of such studies is that
the backreaction of localized fermions on the soliton background is negligible
\cite{Dashen:1974cj,Jackiw:1975fn}, furthermore,
only the fermionic zero modes were considered in most cases.
This assumption is certainly warranted in the weak coupling limit.
However, as the coupling becomes stronger, the effects of the
backreaction can be significant. The problem was revisited recently in
Refs.~\cite{Perapechka:2018yux,Perapechka:2019dvc,Klimashonok:2019iya,Perapechka:2019vqv} where
backreaction of the localized fermions on solitons was taken into account consistently.
This concept was subsequently applied
in investigations of various dynamical systems with localized fermions~\cite{Campos:2022zdp,Weigel:2023fxe,Bazeia:2022yyv,Gani:2022ity,Amari:2024rpm}.

The presence of bounded fermionic modes leads to many interesting features, including
fermion number fractionization \cite{Jackiw:1975fn,Jackiw:1981ee}, string superconductivity
\cite{Witten:1984eb}, and a monopole catalyzed proton decay \cite{Rubakov:1982fp,Callan:1982au}.
A peculiar property of topological solitons is the intrinsic relation between the
the topological degree of the bosonic configuration and the
number of fermionic zero modes localized on a soliton. The Atiyah-Patodi-Singer index theorem~\cite{Atiyah:1975jf}
provides a remarkable relation between these quantities.
Note that, except zero modes, such systems may also support the existence of a tower of
localized fermionic modes with nonzero energy.

However, not much is known about the fermionic modes localized on a non-topological soliton,
like a sphaleron or a Q-ball. Certainly, the index
theorem excludes the existence of fermionic zero modes, however some other mechanisms can exist.
One interesting, less explored option here is to consider an ancillary coupling via electromagnetic interaction,
as suggested in  Refs.~\cite{Anagnostopoulos:2001dh,Levi:2001aw}. Indeed, it is known that Dirac fermions can
be localized by electrostatic interaction \cite{Wakano}. Furthermore, the global $U(1)$ symmetry of Q-balls, which contain a
complex scalar field, can be promoted to a local gauge symmetry \cite{Lee:1988ag,Lee:1991bn,Kusenko:1997vi,Gulamov:2015fya,Gulamov:2013cra,Panin:2016ooo,Nugaev:2019vru,Loiko:2019gwk}.
It is important to note that we consider a single normalizable solutions of the Dirac equation coupled to the electromagnetic field of a gauged Q-ball, and such a
configuration cannot be spherically symmetric.
Furthermore, an axially symmetric Q-ball possesses an intrinsic angular momentum \cite{Volkov:2002aj},
and the corresponding scalar current is a source in the Maxwell equations: it yields both electric and
magnetic fields of the $U(1)$ gauged Q-ball. As we will see, Dirac fermions can be localized in such a field.

It is the
main purpose of this work to
 construct such bounded configurations and investigate their properties.
 This paper is organized as follows. In Sec.~\ref{model},  we
introduce the model which describes $U(1)$ gauged Q-balls with the usual self-interaction potential and Dirac fermions with nonzero bare mass, minimally coupled to the electromagnetic field of the Q-ball.
In Sec.~\ref{ans_eq},  we discuss the axially symmetric parametrization of the fields, the gauge fixing and the physical quantities of interest, like angular momenta and the conserved charges.
The numerical results are presented in Sec.~\ref{num_res}, where we introduce the boundary conditions under which the field equations are solved numerically,
discuss the asymptotic behavior of the fields and analyze some properties of the solutions. Finally, in Sec.~\ref{conclus}, we give our conclusions.

\section{The model}
\label{model}

We are looking for localized configurations consisting of a spinor field $\psi$ minimally coupled to the Maxwell field $A_\mu$ and complex scalar field $\phi$.
The Lagrangian of the system can be written in the form\footnote{We do not include Yukawa interaction term here, as well as possible parity violating terms.}
(we use natural units with $c=\hbar=1$ throughout the paper)
\begin{equation}
\label{Lagr_gen}
    \mathcal{L} = - \frac{1}{4} F_{\mu \nu} F^{\mu \nu}
    + \imath \bar{\psi} \gamma^\mu \psi_{; \mu} -m \bar{\psi} \psi
    + \left( D_\mu \phi \right) ^* D^\mu \phi
    - U \left(  \left| \phi \right|^2 \right)
\, ,
\end{equation}
where $m$ is  a bare mass of the fermion field
and the electromagnetic field tensor is $F_{\mu\nu} = \partial_{ \mu} A_\nu - \partial_\nu A_\mu$. For the scalar field, the asterisk denotes complex conjugation; $\bar \psi$ denotes the Dirac conjugate.
The covariant derivative of the complex scalar field is $D_\mu \phi = \partial_\mu\phi-\imath e A_\mu\phi$, and
the covariant derivative of the spinor field is defined as
$
\psi_{; \mu} =  [\partial_{ \mu} +1/8\, \omega_{a b \mu}\left( \gamma^a  \gamma^b- \gamma^b  \gamma^a\right)-\imath\, e A_\mu]\psi
$ with $\gamma^a$ being the Dirac matrices in the flat space. The gauge coupling constant $e$  describes the minimal interaction between
electromagnetic field and spinor/scalar fields.
 The Dirac matrices in curvilinear coordinates are introduced as $\gamma^\mu = e_a^{\phantom{a} \mu} \gamma^a$,
where
$ e_a^{\phantom{a} \mu}$ is a  tetrad, and $\omega_{a b \mu}$ is the spin connection\footnote{
See, e.g., Ref.~\cite{Lawrie2002}, Eq.~(7.135) for related definitions.
}.
In the above expressions, $\mu,\nu=0,1,2,3$ are spacetime indices and $a,b=0,1,2,3$ are tetrad indices. In what follows, we use the Weyl representation of the Dirac matrices,
$$
\gamma^0 =
     \begin{pmatrix}
        0   &   1 \\
        1   &   0
    \end{pmatrix},\quad
\gamma^k =
     \begin{pmatrix}
        0   &   \sigma^k \\
        -\sigma^k   &   0
    \end{pmatrix},
$$
where $k=1,2,3$ and $\sigma^k$ are the Pauli matrices.

Varying the action with the Lagrangian~\eqref{Lagr_gen} with respect to the spinor/scalar fields and to the vector potential $A_\mu$, we derive the corresponding Dirac, Klein-Gordon,  and Maxwell field equations
\begin{align}
    &
    \frac{1}{\sqrt{-\cal{g}}} \frac {\partial}{\partial x^\mu}
    \left(\sqrt{-\cal{g}}F^{\mu \nu}\right) = e \left(j_\psi^{\nu} +  j_\phi^{\nu} \right),
\label{1_10}\\
    &
    \imath \gamma^\mu \psi_{; \mu} -m\psi
=  0 ,
\label{1_20}\\
     D_\mu D^\mu \phi &
     \equiv \frac{1}{\sqrt{-\cal{g}}}
     \frac{\partial}{\partial x^\mu}
     \left(
            \sqrt{-\cal g}\cal{g}^{\mu \nu}\frac{\partial\phi}{\partial x^\nu}
     \right)
    -2 \imath e A^\mu \frac{\partial\phi}{\partial x^\mu}
    - \imath e \phi \frac{1}{\sqrt{-\cal g}}
    \frac{\partial}{\partial x^\mu}\left(\sqrt{-\cal g}A^\mu\right)
    - e^2 A_\mu A^\mu \phi \nonumber \\
    & =
    -\frac{d U}{d\left| \phi \right|^2} \phi,
\label{1_25}
\end{align}
written in curvilinear coordinates with the metric $\cal{g}_{\mu\nu}$.
Here $\cal g$ is the determinant of the metric.
The system of dynamical equations \re{1_10}-\re{1_25} is invariant under the local $U(1)$ gauge transformations
\be
\phi \to \phi e^{ie \alpha},\quad
\psi \to \psi e^{ie \alpha},\quad A_\mu \to A_\mu +\partial_\mu \alpha ,
\label{gauge}
\ee
where $\alpha$ is a real function of spacetime coordinates.

The potential of a self-interacting scalar field, which admits Q-balls, can be chosen, for example, as follows~\cite{Battye:2000qj,Volkov:2002aj,Kleihaus:2005me}:
\begin{equation}
    U \left(  \left| \phi \right|^2 \right) =
    m_s^2 \left| \phi \right|^2
    - \frac{a}{2} \left| \phi \right|^4
    + \frac{b}{3} \left| \phi \right|^6
\label{U_sf}
\end{equation}
with $m_s$ being the scalar field mass. Here $a$ and $b$ are some  parameters which should be chosen so as to ensure the existence of Q-balls.

The currents of the spinor and scalar fields appearing as sources
in the Maxwell equations~\re{1_10} have the following form:
\begin{align}
    j_\psi^{\nu} = & - \bar{\psi} \gamma^\nu \psi ,
\label{1_30}\\
     j_\phi^{\nu} = & \imath  \left[
        \phi \left( D^\nu \phi \right)^* - \phi^*  D^\nu \phi
     \right] .
\label{1_35}
\end{align}
They are Noether currents, which are conserved
because of invariance of the Lagrangian~\re{Lagr_gen} under the transformations~\re{gauge} with a global phase.

\section{Axially symmetric Ansatz and
dynamical equations}
\label{ans_eq}

\subsection{Ans\"{a}tze and the system of equations}

We work in spherical coordinates with the line element
$$
    ds^2 = dt^2 - dr^2 - r^2\left(d\theta^2 + \sin^2\theta d\varphi^2\right) .
$$

The gauge field is parameterized by an electric and a magnetic potentials
\begin{equation}
A_\mu=\{\chi(r,\theta),0,0,\sigma(r,\theta)\}.
\label{EM_ans}
\end{equation}
Hence, nonvanishing components of the electromagnetic field strength tensor are:
\begin{equation}
    E_r = - \frac{\partial \chi}{\partial r}, \quad
    E_\theta = - \frac{\partial \chi}{\partial \theta}, \quad
    H_r = - \frac{\csc\theta}{r^2} \frac{\partial \sigma}{\partial \theta}, \quad
    H_\theta = \csc\theta \frac{\partial \sigma}{\partial r} .\nonumber
\label{EM_components}
\end{equation}

The spinor field is parametrized by two complex functions
\cite{Herdeiro:2019mbz,Herdeiro:2021jgc,Dzhunushaliev:2023vxy}
 \begin{equation}
    \psi^T = e^{\imath \left(M_\psi\varphi-\Omega t\right)}
        \begin{pmatrix}
            \psi_1, & \psi_2, & \psi_2^*, & \psi_1^*
        \end{pmatrix}\, .
\label{spinor}
\end{equation}
 Here $\Omega$ is the eigenvalue of the Dirac Hamiltonian, $M_\psi$ is a half-integer parameter (the azimuthal number).
For our purposes, it is convenient to represent the components of the spinor field \re{spinor} as
$$
    \psi_1=\frac{1}{2}\left[X+Y+\imath\left(V+W\right)\right],\quad
    \psi_2=\frac{1}{2}\left[X-Y+\imath\left(V-W\right)\right],
$$
where the four real functions $X,Y,V$, and $W$ depend only on the spherical coordinates $r$ and $\theta$.

Imposing axial symmetry, we employ for the scalar field the usual Ansatz  \cite{Volkov:2002aj,Kleihaus:2005me}
\begin{equation}
\phi=\Phi(r,\theta) e^{\imath\left( n \varphi+\omega t\right)},
\label{sf_ans}
\end{equation}
where $\omega$ is the scalar field angular frequency, which we shall take to be positive,
and $n$ is an integer parameter (rotational number).

Note that invariance of the model with respect to the Abelian gauge transformations~\re{gauge} allows us to adjust the asymptotic value
of the electric potential $V\equiv \chi(\infty)$ (this symbol $V$ should not be confused with that used for the spinor field)
and the angular frequency as $V \to V +\frac{1}{e}\xi,~~ \omega \to \omega + \xi$. In particular, setting  $\xi=-eV$ yields $\chi(\infty)=0$. In such a stationary gauge
the scalar field obtains an explicit time dependence with angular frequency~$\omega$ associated with time-dependent gauge transformations~\cite{Radu:2005jp,Kleihaus:2009kr,Herdeiro:2020xmb,Loginov:2020xoj}.

Alternatively, one can perform a gauge transformation
to absorb both the azimuthal  and temporal dependencies  of the scalar field into the
transformed electromagnetic potential~\cite{Herdeiro:2021jgc}.
Here we shall follow the related discussion of Refs.~\cite{Loiko:2019gwk,Herdeiro:2021jgc,Kunz:2024uux,Herdeiro:2024yqa,Kirichenkov:2023omy}  and employ the stationary gauge fixing
$\chi(\infty)=0$ and retaining the dependence of the solutions on the angular frequency $\omega$.

Substitution of the Ans\"{a}tze  \eqref{EM_ans}, \eqref{spinor}, and \eqref{sf_ans}  into the field equations~\eqref{1_10}-\eqref{1_25}
leads to a system of seven coupled partial differential equations \eqref{Dirac_eq_X}-\eqref{sf_eq_phi} that have to be solved numerically subject to appropriate boundary conditions.
In order to simplify the consideration, hereafter we restrict ourselves to the case of axially symmetric configurations with
$M_\psi=1/2$ and $n=1$. Further, the mass of the scalar excitations $m_s$ provides a characteristic scale, which can be used to introduce new dimensionless variables: $x=m_s r$, $\tilde\Omega,\tilde\omega=\Omega,\omega/m_s$,
$(\tilde X, \tilde Y, \tilde V, \tilde W)=(X,Y,V,W)/m_s^{3/2}$,
$\tilde{m}=m/m_s$,
$\tilde{\chi}=\chi/m_s$,
$\tilde{\Phi}=\Phi/m_s$, $\tilde b=m_s^2 b$.
Thus, the resulting system depends on a set of input parameters  $\tilde\omega, \tilde m, a, \tilde{b}$, and $e$.
In turn, the eigenvalue of the Dirac Hamiltonian $\tilde \Omega$ is not a free parameter,
and it has to be determined so as to obtain a particular normalizable solution, see Ref.~\cite{Perapechka:2018yux}.
Further, the parameters of the scalar potential $a$ and $\tilde b$ should be taken so as to ensure the existence of a Q-ball.
Taking all this into account, the field equations of the model become
\begin{align}
    \tilde X_{,x} + \frac{\tilde X}{x} - \frac{\tilde W_{,\theta}}{x}
  - \frac{\cot\frac{\theta}{2}}{2x} \tilde W
  + e
  \left(
     \frac{\csc\theta }{x}\tilde W  \sigma
   - \tilde V \tilde \chi
   \right) - \left(\tilde{m} + \tilde \Omega\right)\tilde V
= 0 ,
\label{Dirac_eq_X}\\
  \tilde Y_{,x} + \frac{\tilde Y}{x} - \frac{\tilde V_{,\theta}}{x}
   + \frac{\tan\frac{\theta}{2}}{2x} \tilde V
   + e
   \left(
        -\frac{\csc\theta }{x}\tilde V  \sigma + \tilde W \tilde \chi
    \right) - \left( \tilde{m} - \tilde \Omega\right)\tilde W
= 0 ,
\label{Dirac_eq_Y}\\
   \tilde V_{,x} + \frac{\tilde V}{x} + \frac{\tilde Y_{,\theta}}{x}
    + \frac{\cot\frac{\theta}{2}}{2x}\tilde Y
    + e
    \left(
        -\frac{\csc\theta }{x}\tilde Y  \sigma+\tilde X \tilde \chi
    \right) - \left(\tilde{m} - \tilde \Omega\right) \tilde X
= 0 ,
\label{Dirac_eq_V}\\
  \tilde W_{,x} + \frac{\tilde W}{x} + \frac{\tilde X_{,\theta}}{x}
   - \frac{\tan\frac{\theta}{2}}{2x} \tilde X
   + e
   \left(
     \frac{\csc\theta }{x}\tilde X \sigma - \tilde Y \tilde \chi
   \right)-\left(\tilde{m}+\tilde \Omega\right)\tilde Y
= 0 ,
\label{Dirac_eq_W}\\
  \tilde \chi_{,xx} + \frac{2}{x}\tilde \chi_{,x} +
   \frac{1}{x^2}\tilde \chi_{,\theta\theta}
   + \frac{\cot\theta}{x^2}\tilde \chi_{,\theta} +2 e\left(\tilde\omega-e\tilde\chi\right)\tilde{\Phi}^2
   - e\, U_1  = 0 ,
\label{Maxw_eq_phi}\\
  \sigma_{,xx} + \frac{1}{x^2} \sigma_{,\theta\theta}
  - \frac{\cot\theta}{x^2} \sigma_{,\theta} +2 e\left(1-e\sigma\right)\tilde{\Phi}^2
  - 2\, e\, x \sin\theta\, U_2  = 0 ,
\label{Maxw_eq_sigma}
\end{align}
\begin{align}
  \tilde \Phi_{,xx}+\frac{2}{x}\tilde \Phi_{,x}  + \frac{1}{x^2} \tilde\Phi_{,\theta\theta}
  + \frac{\cot\theta}{x^2} \tilde\Phi_{,\theta}
  +e^2\left(\tilde\chi^2-\frac{\csc^2\theta}{x^2}\sigma^2\right)\tilde\Phi+
  2 e\left(-\tilde\omega\tilde\chi+\frac{\csc^2\theta}{x^2}\sigma\right)\tilde\Phi
\nonumber\\
    - \left(1-a\tilde\Phi^2
    + \tilde b\tilde\Phi^4\right)\tilde{\Phi}
    + \left(\tilde\omega^2
    - \frac{\csc^2\theta}{x^2}\right)\tilde\Phi
  = 0 ,
\label{sf_eq_phi}
\end{align}
where
$$
    U_1 = \tilde X^2 + \tilde Y^2 + \tilde V^2 + \tilde W^2, \quad
    U_2 = \tilde X \tilde Y + \tilde V \tilde W
$$
and lower indices denote differentiation with respect to the corresponding coordinates. Notice that these equations  are invariant with respect to multiplying the spinor functions by~$-1$.

This system is supplemented by the normalization condition imposed on the spinor field
\be
\int d^3 x\, \psi^\dag \psi =\int d^3 x \left(X^2+Y^2+V^2+W^2\right) =1 .
\label{norm}
\ee
In other words, we restrict our consideration to a single-particle fermion state. In our numerical scheme we explicitly
impose this condition.

\subsection{Quantities of interest}

Let us now write down expressions for some physically interesting quantities which will be
useful for understanding the properties of the configurations in question.

From the Lagrangian~\eqref{Lagr_gen}, one can obtain the corresponding energy-momentum tensor of the system under consideration (already in a symmetric form)
\begin{align}
\label{EM}
    T_{\mu}^\nu &= \frac{\imath }{4}g^{\nu\rho}
    \left[
        \bar\psi \gamma_{\mu} \psi_{;\rho}
        + \bar\psi\gamma_\rho\psi_{;\mu}
    - \bar\psi_{;\mu}\gamma_{\rho }\psi
    - \bar\psi_{;\rho}\gamma_\mu\psi
    \right]
    - F^{\nu\rho} F_{\mu\rho}
    + \frac{1}{4} \delta_\mu^\nu F_{\alpha\beta} F^{\alpha\beta}
\nonumber\\
    & + {\cal g}^{\nu\sigma}\left[\left( D_\sigma \phi \right) ^* D_\mu \phi
    + D_\sigma \phi \left(D_\mu \phi \right) ^*\right]
\\
    & - \delta^\nu_\mu\left\{
        \frac{1}{2}{\cal g}^{\lambda\sigma}
        \left[
            \left( D_\lambda \phi \right) ^* D_\sigma \phi+ \left(D_\sigma \phi \right) ^* D_\lambda \phi
        \right]-
        U \left( \left| \phi \right|^2 \right)
    \right\}.
\nonumber
\end{align}

The total dimensionless mass of the system can be found in the form
\begin{equation}
\tilde{M}\equiv  M/m_s = 2 \pi \int_0^\infty dx \int_0^\pi  d\theta\, \tilde{T}_t^t x^2 \sin\theta   ,
\label{M_tot}
\end{equation}
where the dimensionless $(^t_t)$-component of the energy-momentum tensor~\eqref{EM} is
\begin{equation}
\begin{split}
    \tilde{T}_t^t\equiv T_t^t/m_s^4 &= \frac{1}{2}\left[
        \tilde \chi_{,x}^2 + \frac{\csc^2\theta }{x^2} \sigma_{,x}^2
        + \frac{1}{x^2}\tilde \chi_{,\theta}^2
        + \frac{\csc^2\theta }{x^4} \sigma_{,\theta}^2
        + 2 \left(\tilde \Omega + e \tilde \Phi\right) U_1
        + U_3^2
    \right]  \\
&+\tilde \Phi_{,x}^2+ \frac{1}{x^2}\tilde \Phi_{,\theta}^2 + e^2\left(\frac{\csc^2\theta}{x^2}\sigma^2+\tilde\chi^2\right)\tilde\Phi^2
-2 e\left(\frac{ \csc^2\theta }{x^2}\sigma+\tilde\omega\tilde\chi\right)\tilde\Phi^2
\\
&+\frac{\csc^2\theta}{x^2}\tilde\Phi^2+\left(\tilde\omega^2+1-\frac{a}{2}\tilde\Phi^2+\frac{\tilde b}{3}\tilde\Phi^4\right)\tilde\Phi^2
\label{T_tt}
\end{split}
\end{equation}
with $U_3 = \tilde X^2 - \tilde Y^2 - \tilde V^2 + \tilde W^2$.

The total  dimensionless angular momentum reads
\begin{equation}
 \tilde J =2\pi \int_{0}^{\infty} dx \int_{0}^{\pi} d\theta\, \tilde T_\varphi^t x^2 \sin\theta  ,
\label{ang_mom_tot}
\end{equation}
where the dimensionless $(^t_\varphi)$-component of the energy-momentum tensor~\eqref{EM} is
\begin{equation}
\begin{split}
    \tilde{T}_\varphi^t\equiv T_\varphi^t/m_s^3 &=  \tilde \chi_{,x} \sigma_{,x}
    + \frac{1}{x^2} \tilde \chi_{,\theta} \sigma_{,\theta}
    - \frac{1}{4}\left(1-2\, e\,  \sigma\right) U_1
    + x\sin\theta \left(\tilde \Omega
    + e \tilde \chi\right) U_2
\\
    & + \frac{1}{2}\sin\theta\left(
        \tilde W \tilde X - \tilde V \tilde Y
    \right)
    + \frac{1}{4} \cos\theta
    \left(
        \tilde X^2-\tilde Y^2+\tilde V^2-\tilde W^2
    \right)
\\
    & + 2\left[
         \tilde\omega
        + e^2\tilde\chi\sigma
        - e\left( \tilde\chi + \tilde\omega\sigma\right)
    \right]\tilde \Phi^2 .
\label{T_tphi}
\end{split}
\end{equation}
The occurrence of a nonzero angular momentum \re{ang_mom_tot} is due to the presence in the system of (i)~a single fermion possessing an intrinsic  half-integer momentum;
 (ii)~crossed electric and magnetic fields; and (iii)~an internally rotating axially symmetric Q-ball.

The conserved scalar charge $Q_\phi$ is defined as
the integral of the temporal component of
the  four-current~\eqref{1_35},
\begin{equation}
Q_\phi=
\int  j^t_\phi \sqrt{{-\cal g}}\,dr d\theta d\varphi
=4\pi \int_0^\infty dr \int_{0}^{\pi}d\theta \,\tilde{\Phi}^2\left(\tilde\omega-e \tilde\chi\right)x^2 \sin\theta \, . \nonumber
\label{scalar_charge}
\end{equation}
Further, the spinor charge associated with the  Noether current is obtained using
Eq.~\eqref{1_30} in the form
\begin{equation}
Q_\psi=  \int  j^t_\psi \sqrt{{-\cal g}}\,dr d\theta d\varphi  = 
    2 \pi \int_{0}^{\infty}dx\int_{0}^{\pi}d\theta
    \left(
        \tilde X^2 + \tilde Y^2 + \tilde V^2 + \tilde W^2
    \right) x^2 \sin\theta  . \nonumber
\label{spinor_charge}
\end{equation}
Clearly, the normalization condition \re{norm} implies that for a localized fermionic mode $Q_\psi =1$.

Note that there is a general relation for the angular momentum of a Q-ball, $J_\phi$, and the scalar charge $Q_\phi$: $J_\phi=n Q_\phi$ \cite{Volkov:2002aj,Kleihaus:2005me}.
Similar  relation also holds for the spinor field \cite{Herdeiro:2019mbz}, leading, in the general case, to
$$
J\equiv J_\phi+J_\psi=n Q_\phi- M_\psi Q_\psi .
$$
Consequently, the solutions can be thought of as corresponding to minima of the total energy functional \re{M_tot}
with fixed angular momentum.
It should be noted that this relation holds for the
integrated components of the total angular momentum, although the corresponding densities may be different
\cite{Herdeiro:2019mbz}. Our numerical results
confirm this relation\footnote{Recall that we restrict our considerations to the case $n=1$, $M_\psi=1/2$.}.

\section{Numerical results}
\label{num_res}

\subsection{Boundary conditions and the numerical approach}

The choice of appropriate boundary conditions must guarantee that the localized axially symmetric solutions
of the system of seven partial differential equations~\eqref{Dirac_eq_X}-\eqref{sf_eq_phi}
are asymptotically flat, globally regular, and that they possess finite energy. The corresponding expansions of these equations at the origin, on the spatial boundary and on the symmetry axis, yield the following  requirements:
\begin{equation}
\begin{split}
    \left. \tilde\Phi \right|_{x = 0}=0, \left. \frac{\partial  \tilde X}{\partial x}\right|_{x = 0} = \left. \frac{\partial \tilde Y}{\partial x}\right|_{x = 0}=\left. \frac{\partial  \tilde V}{\partial x}\right|_{x = 0}=
    \left. \frac{\partial \tilde W}{\partial x}\right|_{x = 0} =
  \left. \frac{\partial \tilde \chi}{\partial x}\right|_{x = 0}=  0,  \left.  \sigma \right|_{x = 0}= 0;
\\
\left. \tilde \Phi \right|_{x = \infty} =\left. \tilde X \right|_{x = \infty} =
    \left. \tilde Y \right|_{x = \infty} =
  \left. \tilde V \right|_{x = \infty} =
  \left. \tilde W \right|_{x = \infty} =
    \left.  \tilde\chi \right|_{x = \infty} =
  \left.  \sigma \right|_{x = \infty} =   0 ;
\\
    \left. \tilde\Phi \right|_{\theta = 0} =0, \left. \frac{\partial \tilde X}{\partial \theta}\right|_{\theta = 0} =
  \left. \frac{\partial \tilde V}{\partial \theta}\right|_{\theta = 0} =
    \left. \frac{\partial  \tilde\chi}{\partial \theta}\right|_{\theta = 0} =  0 ,  \left. \tilde Y \right|_{\theta = 0} =\left. \tilde W \right|_{\theta = 0}=\left.  \sigma \right|_{\theta = 0}= 0 ;
\\
\left. \tilde\Phi \right|_{\theta = \pi} =0,  \left. \frac{\partial \tilde Y}{\partial \theta}\right|_{\theta = \pi} =
    \left. \frac{\partial \tilde W}{\partial \theta}\right|_{\theta = \pi} =
    \left. \frac{\partial \tilde \chi}{\partial \theta}\right|_{\theta = \pi} =  0 ,  \left.  \tilde X \right|_{\theta = \pi} =\left.  \tilde V \right|_{\theta = \pi}=\left.  \sigma \right|_{\theta = \pi}= 0.
\label{BCs_tot}
\end{split}
\nonumber
\end{equation}
In our numerical calculations we introduce a
new compact radial coordinate
\begin{equation}
    \bar x = \frac{x}{c_k+x}\, ,
\label{comp_coord}
\end{equation}
which maps the semi-infinite region $[0;\infty)$ onto the unit interval $[0; 1]$. Here $c_k$ is an arbitrary constant which is used to adjust the contraction of
the grid. In our calculations, we typically take $c_k\in [1,20]$.
The emerging system of nonlinear algebraic equations is solved using a trust-region
Newton method \cite{Lin}.
The underlying linear system is solved with the Intel MKL
PARDISO sparse direct solver \cite{pardiso} equipped with an adaptive mesh selection procedure, and the CESDSOL library\footnote{Complex Equations-Simple Domain partial differential equations SOLver, a C++ package developed by I.~Perapechka, see, e.g., Refs. \cite{Kunz:2019sgn,Herdeiro:2019mbz,Herdeiro:2021jgc,Dzhunushaliev:2023ylf,Dzhunushaliev:2024iag}.}.
 Typical mesh sizes include $[200,1000]\times 100$  points covering  the integration region $0\leq \bar x \leq 1$ [given by the compact radial coordinate~\eqref{comp_coord}] and $0\leq \theta \leq \pi$. In all cases, the typical errors are of order of $10^{-4}$.

\subsection{Asymptotic behavior}

The far field asymptotic of the Maxwell equations~\eqref{Maxw_eq_phi} and \eqref{Maxw_eq_sigma} (as $x\to \infty$) is of the form:
\begin{equation}
    \tilde \chi \approx \frac{1}{4\pi} \frac{q_e}{x} + \dots, \quad
    \sigma \approx - \frac{1}{4\pi}\frac{\tilde{\mu}_m}{x} \sin^2 \theta +\dots ,
\label{asympt_behav}
\end{equation}
where $q_e$ is an electric charge of the scalar field and $\tilde{\mu}_m\equiv m_s \mu_m$  is a corresponding magnetic moment. Consequently,  these quantities can be read off from asymptotic subleading behavior of~\eqref{asympt_behav} as
\begin{equation}
    q_e = - 4 \pi \lim_{x\to\infty} x^2 \frac{\partial\tilde \chi}{\partial x}
    = - 4 \pi \lim_{\bar{x}\to 1} \bar{x}^2 \frac{\partial\tilde \chi}{\partial\bar{x}}, \quad
    \tilde{\mu}_m = 4 \pi \lim_{x\to\infty} \frac{x^2}{\sin^2\theta}
    \frac{\partial \sigma}{\partial x}
    = 4 \pi \lim_{\bar{x}\to 1} \frac{\bar{x}^2}{\sin^2\theta} \frac{\partial \sigma}{\partial\bar{x}} . \nonumber
\label{asympt_charge}
\end{equation}

In turn, the asymptotic behavior of the spinor fields follows from the Dirac equations ~\eqref{Dirac_eq_X}-\eqref{Dirac_eq_W},
\be
\begin{split}
\tilde X & \approx -2 \cos\frac{\theta}{2}\, g(x)+\dots, \quad
    \tilde Y\approx 2 \sin\frac{\theta}{2}\, f(x)+\dots, \\
\tilde V & \approx 2 \cos\frac{\theta}{2}\, f(x) +\dots, \quad
    \tilde W\approx -2 \sin\frac{\theta}{2}\, g(x) +\dots. \nonumber
\end{split}
\ee
The explicit form of the functions $f(x)$ and $g(x)$ which appear here depends on the eigenvalues of the Dirac Hamiltonian $\tilde{\Omega}$.
For~$\tilde{\Omega}<\tilde m$, we find
\begin{equation}
    f(x) \approx f_\infty \frac{e^{-\sqrt{\tilde m^2 - \tilde\Omega^2}\,x}}{x}
    x^{-\frac{1}{\sqrt{\tilde m^2 - \tilde\Omega^2}}\frac{e q_e}{4\pi}}, \quad
    g(x) \approx f_\infty \sqrt{\frac{\tilde m + \tilde\Omega}{\tilde m
    - \tilde\Omega}}\,
    \frac{e^{-\sqrt{\tilde m^2 - \tilde\Omega^2}\,x}}{x}
    x^{-\frac{1}{\sqrt{\tilde m^2-\tilde\Omega^2}}\frac{e q_e}{4\pi}}, \nonumber
\label{asympt}
\end{equation}
where $f_\infty$ is an integration constant. In the case of $\tilde{\Omega}=\tilde m$, we have
$$
    f(x) \approx f_\infty \frac{e^{-\sqrt{\frac{2}{\pi} e q_e\,x}}}{x^{5/4}}, \quad
    g(x) \approx f_\infty \sqrt{\frac{8\pi}{e q_e}}\,
    \frac{e^{-\sqrt{\frac{2}{\pi} e q_e\,x}}}{x^{3/4}} .
$$
We notice that regular localized solutions on the threshold
$\tilde{\Omega}=\tilde m$  can exist only for nonvanishing electric charge $q_e$.

Finally, the asymptotic behavior of the scalar field, which follows from Eq.~\eqref{sf_eq_phi}, is of the form:
\begin{align}
&\text{when} \quad \tilde{\omega}<1: \quad \tilde{\Phi}\approx \Phi_\infty e^{-\sqrt{1-\tilde\omega^2}\,x}x^{-\left(1+\frac{ e q_e \tilde\omega}{4\pi\sqrt{1-\tilde\omega^2}}\right)};\nonumber\\
&\text{when} \quad \tilde{\omega}=1: \quad \tilde{\Phi}\approx \Phi_\infty \frac{e^{-\sqrt{\frac{2}{\pi}e q_e \tilde\omega \,x}}}{x^{3/4}},\nonumber
\end{align}
where $\Phi_\infty$ is an integration constant.
Hence, we may conclude that localized solutions of the system \eqref{Dirac_eq_X}-\eqref{sf_eq_phi} can exist if $\tilde\omega < 1$ (both in the presence of the charge $q_e$ and without it) and if $\tilde\omega = 1$ but only when $q_e\neq 0$.

\subsection{Numerical solutions}

\begin{figure}[t]
    \begin{center}
        \includegraphics[width=0.49\linewidth]{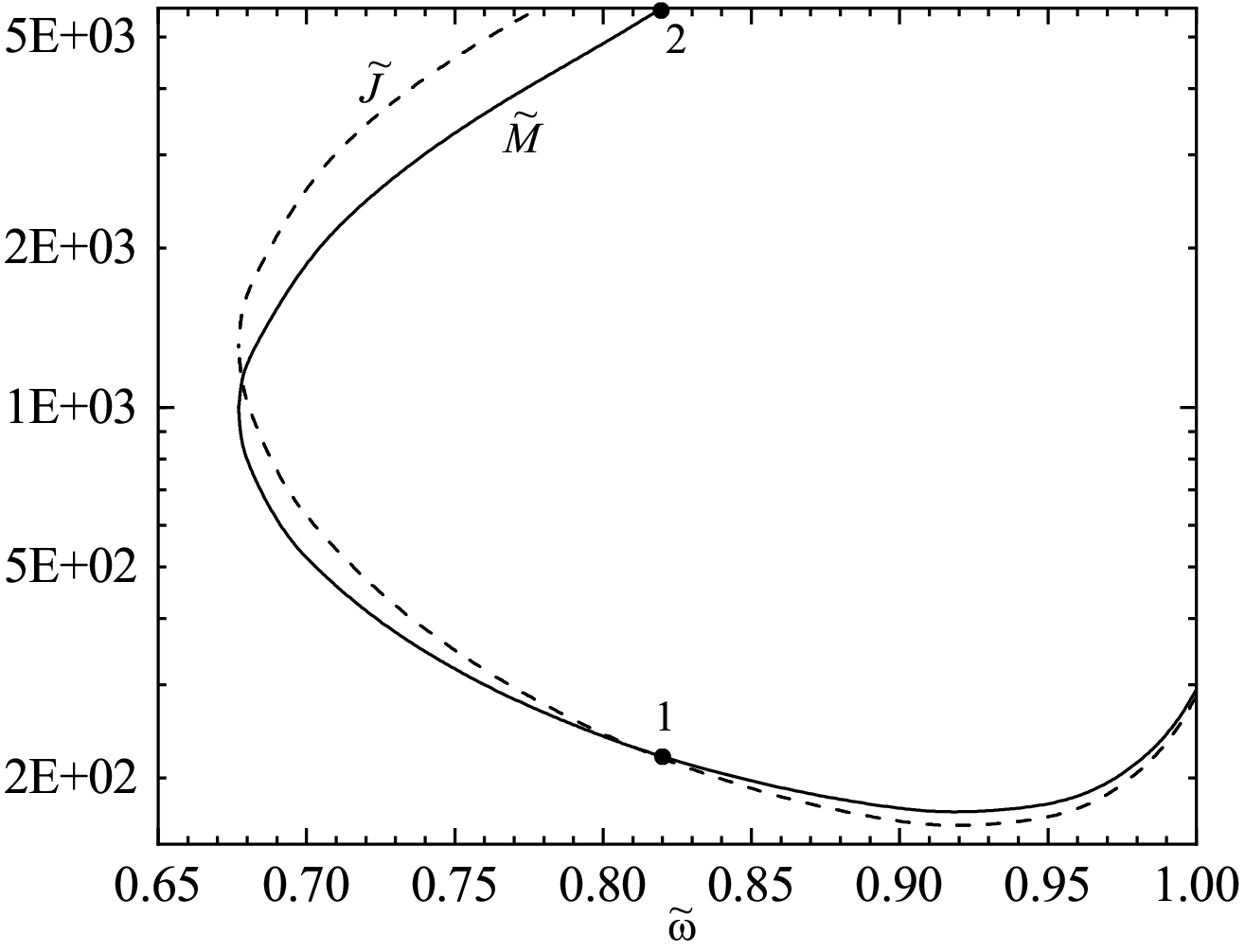}
        \includegraphics[width=0.46\linewidth]{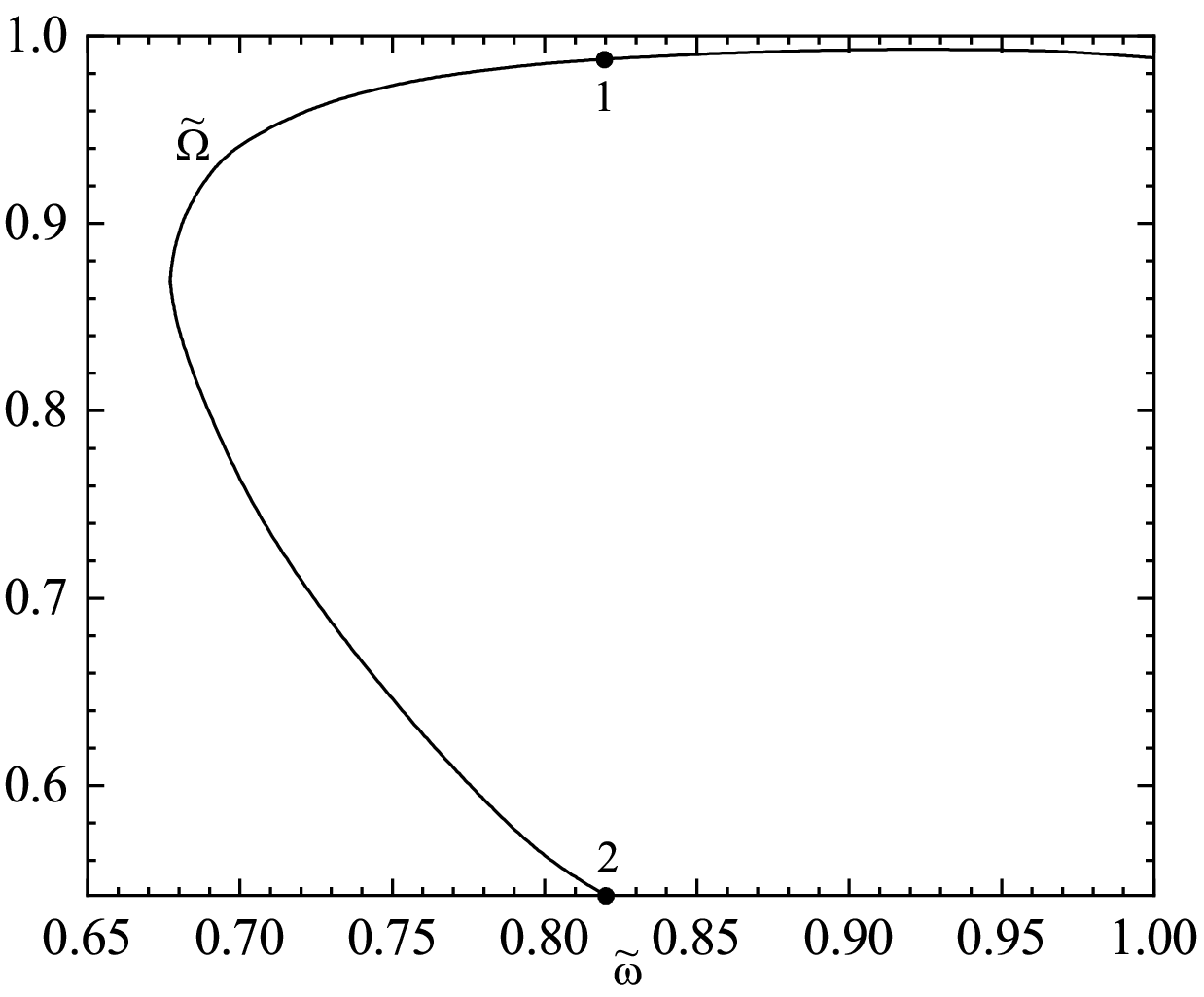}
        \includegraphics[width=0.49\linewidth]{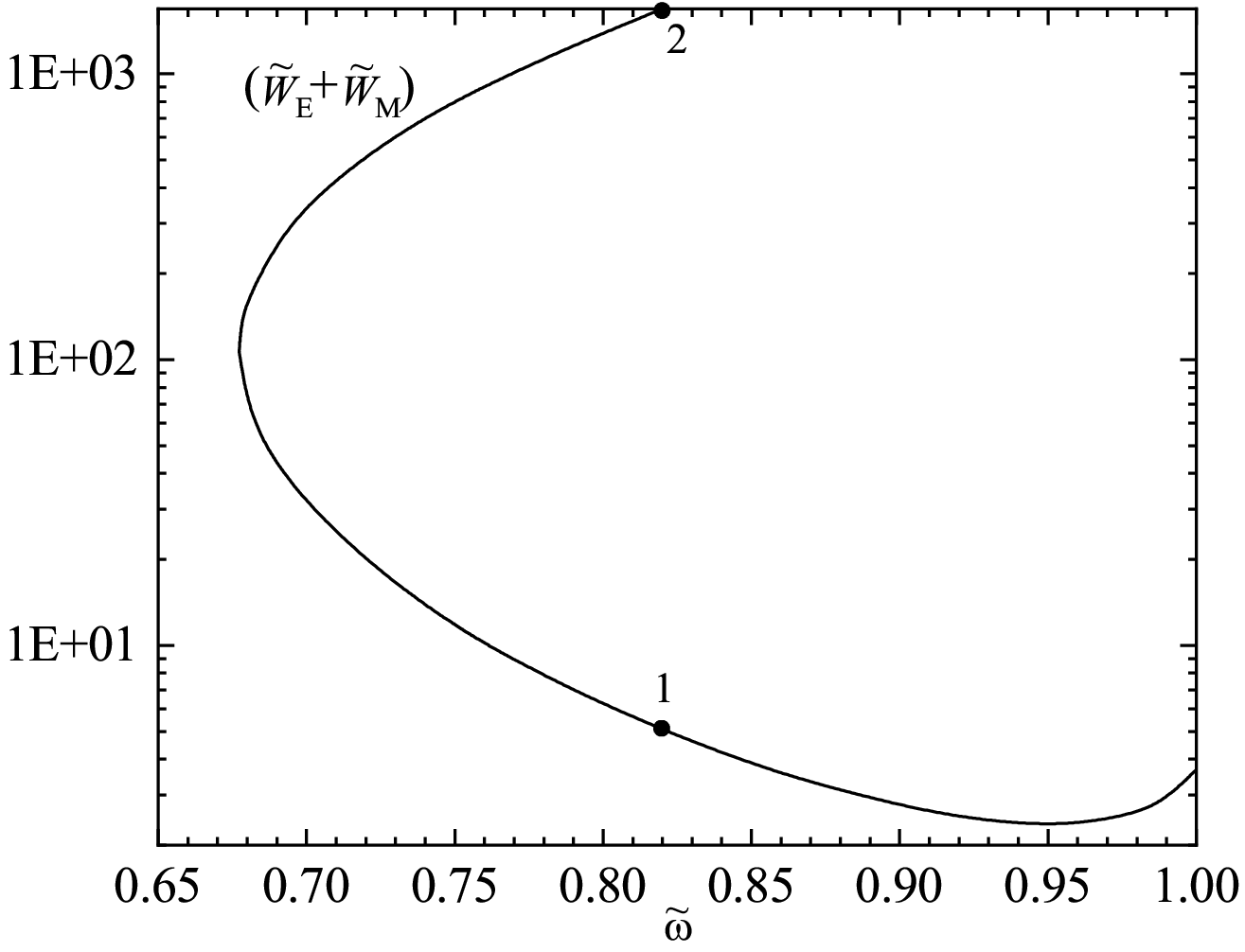}
        \includegraphics[width=0.46\linewidth]{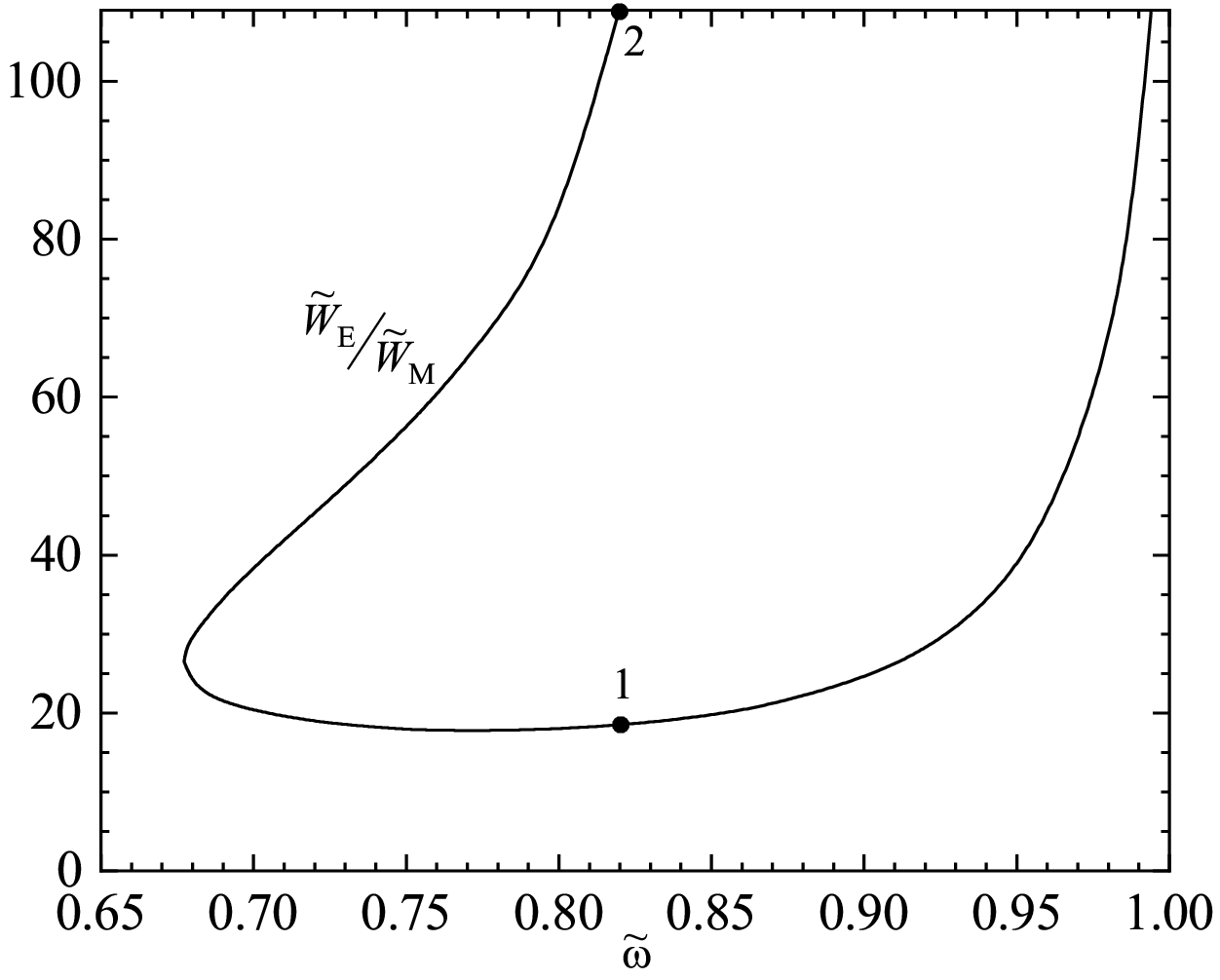}
    \end{center}
    \vspace{-1.cm}
    \caption{The total mass of the system  $\tilde M$ and the total angular momentum $\tilde J$ (upper left panel),
    the eigenvalues of the Dirac Hamiltonian $\tilde \Omega$ (upper right panel), the sum (lower left panel) and the ratio (lower right panel)
    of the electric, $\tilde W_{\text{E}}$, and magnetic, $\tilde W_{\text{M}}$, field energies
    vs. the field frequency $\tilde \omega$.
    The parameters of the potential~\eqref{U_sf} are fixed as $a=3.8, \tilde{b}=3$, the gauge coupling constant is $e=0.1$, and
    the bare mass of the fermion is $\tilde{m}=1$.
    The dots 1 and 2 correspond to the particular solutions
displayed in Fig.~\ref{fig_fields_distr}.}
    \label{fig_M_freq_sf_sp}
\end{figure}

\begin{figure}[h!]
    \begin{center}
        \includegraphics[width=.64\linewidth]{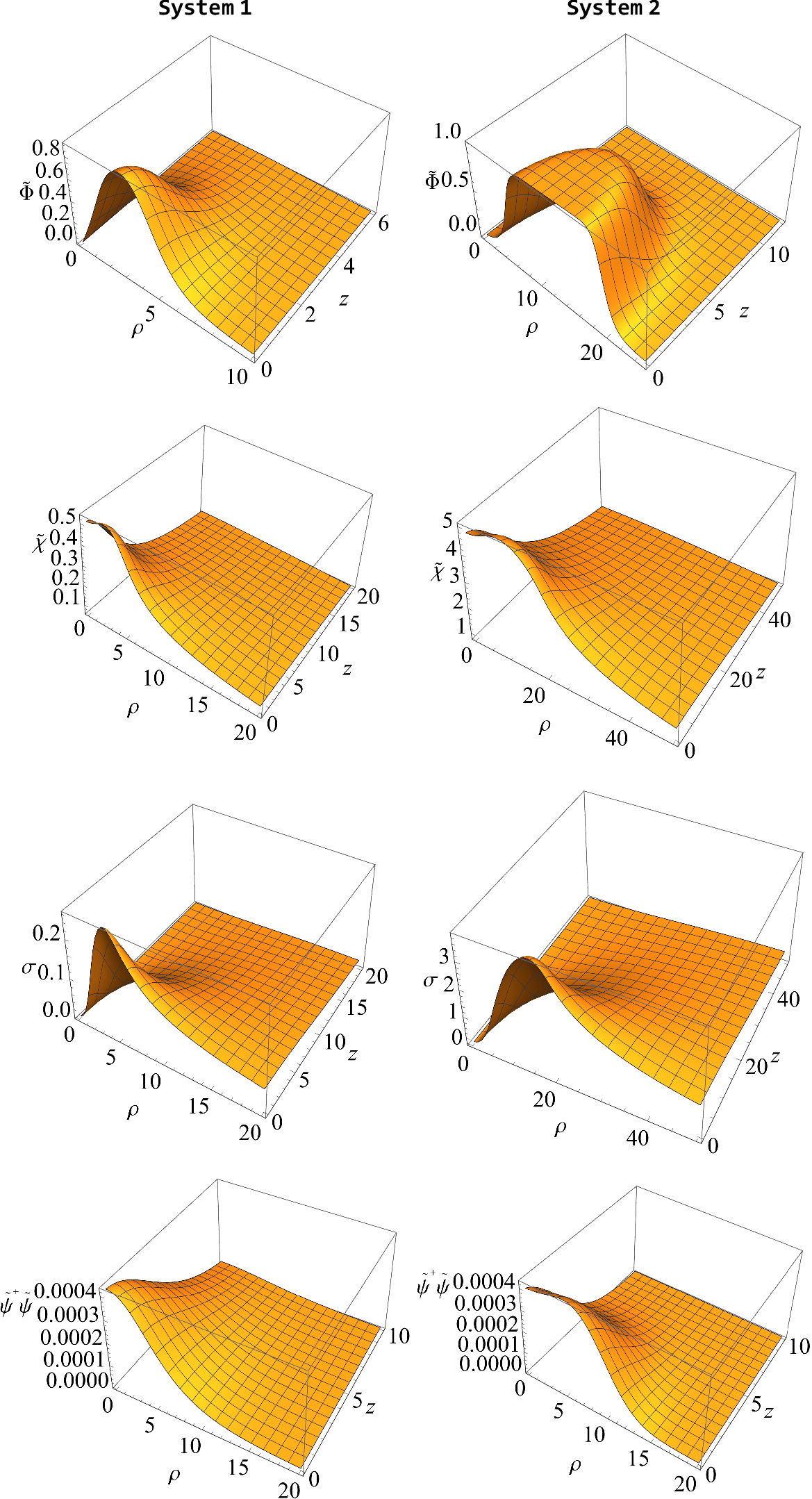}
     \end{center}
    \vspace{-1.cm}
    \caption{Gauged Q-balls with localized fermions on the first branch (left column) and on the second branch (right column).
    The  distributions of the scalar field $\tilde \Phi$, electric potential $\tilde \chi$, magnetic potential $\sigma$ and fermionic density
    $\tilde{\psi}^\dag \tilde \psi$, from top to bottom, are shown for the solutions labeled by the dots 1 and 2 on the curves in Fig.~\ref{fig_M_freq_sf_sp} as
    functions of the coordinates $\rho=x \sin\theta$ and $z=x \cos\theta$.}
    \label{fig_fields_distr}
\end{figure}

\begin{figure}[t]
    \begin{center}
        \includegraphics[width=.543\linewidth]{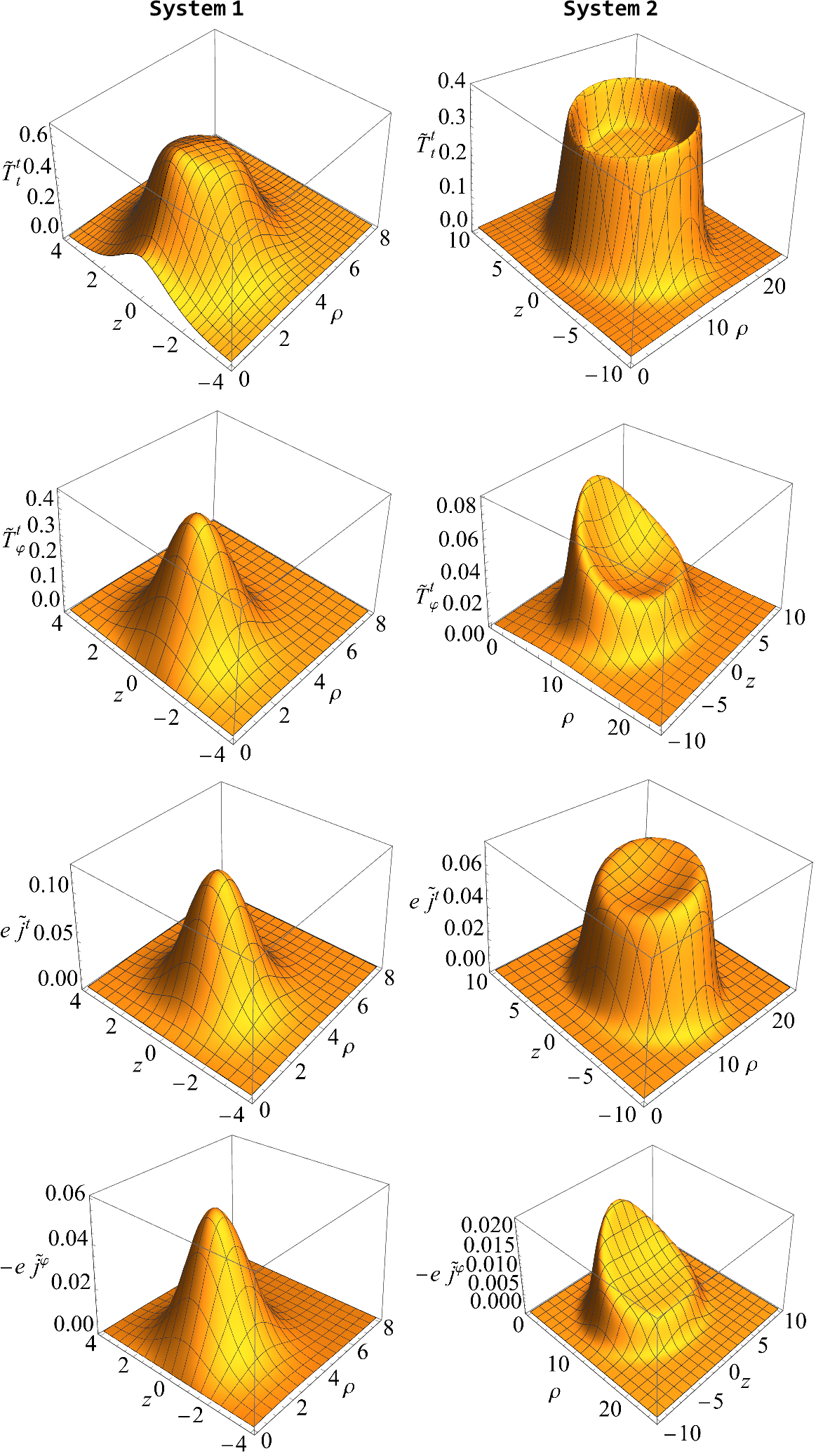}
     \end{center}
    \vspace{-1.cm}
    \caption{Gauged Q-balls with localized fermions on the first branch (left column) and on the second branch (right column).
    The  distributions of the
    the dimensionless energy density~$\tilde T_t^t$ from Eq.~\eqref{T_tt}, the physical component of the angular momentum density $\tilde T_\varphi^t$ from Eq.~\eqref{T_tphi},
    the total charge density $\tilde{j}^t= \tilde{j}^t_\psi+\tilde{j}^t_\phi$ and physical component of the total current density
$\tilde{j}^\varphi=\tilde{j}^\varphi_\psi+\tilde{j}^\varphi_\phi$, from top to bottom, are shown for the solutions labeled by the dots 1 and 2 on the curves in Fig.~\ref{fig_M_freq_sf_sp} as
    functions of the coordinates $\rho=x \sin\theta$ and $z=x \cos\theta$. }
    \label{fig_EMT_currents}
\end{figure}

\begin{figure}[t]
    \begin{center}
        \includegraphics[width=1.\linewidth]{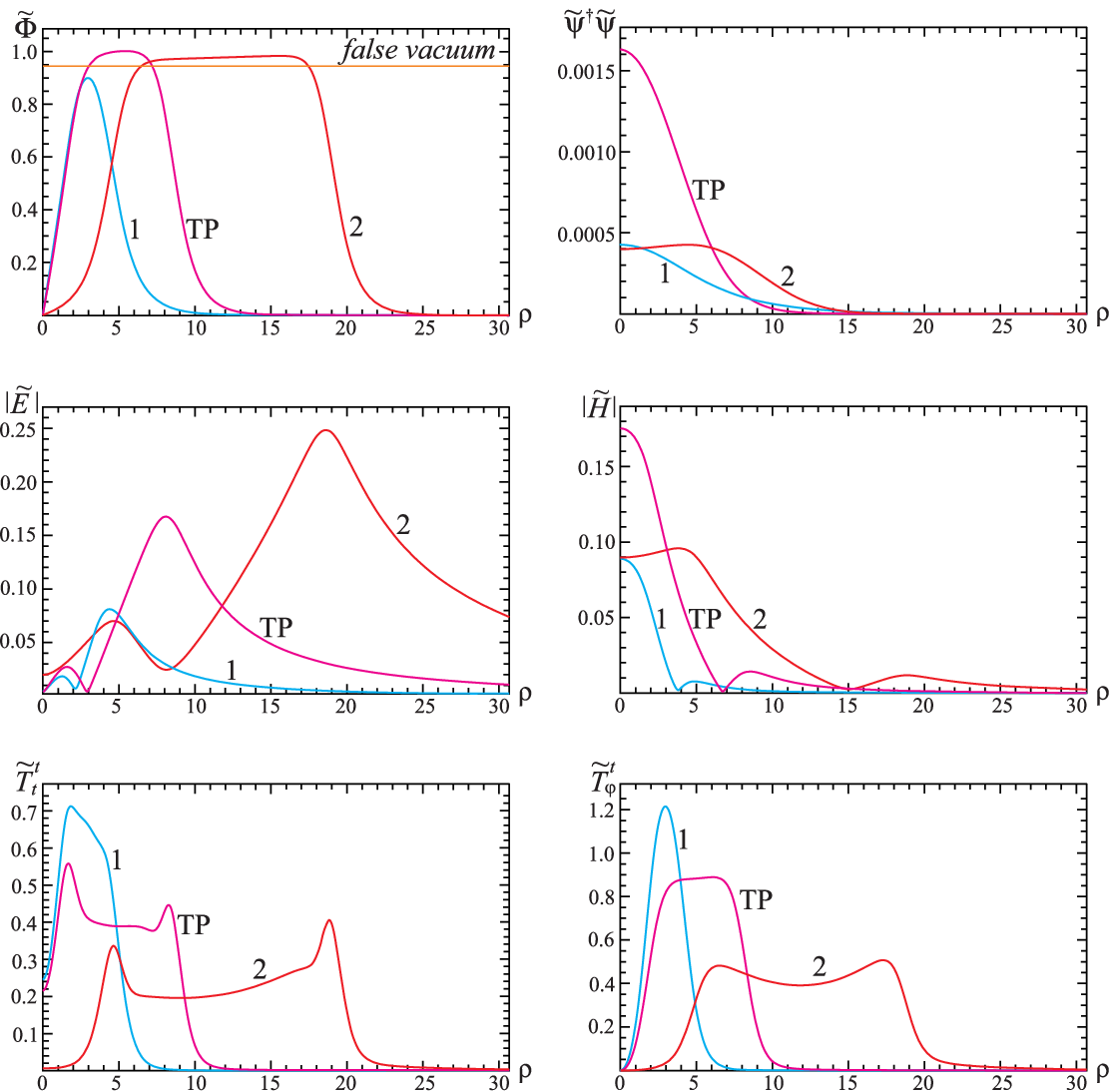}
     \end{center}
    \vspace{-.5cm}
    \caption{The radial distributions of the scalar field $\tilde{\Phi}$, spinor field density $\tilde{\psi}^\dag \tilde{\psi}$, modulus of the electric, $|\tilde{E}|$, and magnetic, $|\tilde{H}|$, field strengths,
     energy density $\tilde{T}^t_t$, and angular momentum density~$\tilde{T}^t_\varphi$
    in the equatorial plane ($\theta=\pi/2$) for the configurations labeled by the dots 1 and 2 in Fig.~\ref{fig_M_freq_sf_sp},
    as well as for the configuration lying at the turning point (TP) where $\tilde{\omega}=\tilde{\omega}_{\text{min}}$.
     }
    \label{fig_fields_distr_equator}
\end{figure}

\begin{figure}[t]
    \begin{center}
        \includegraphics[width=.85\linewidth]{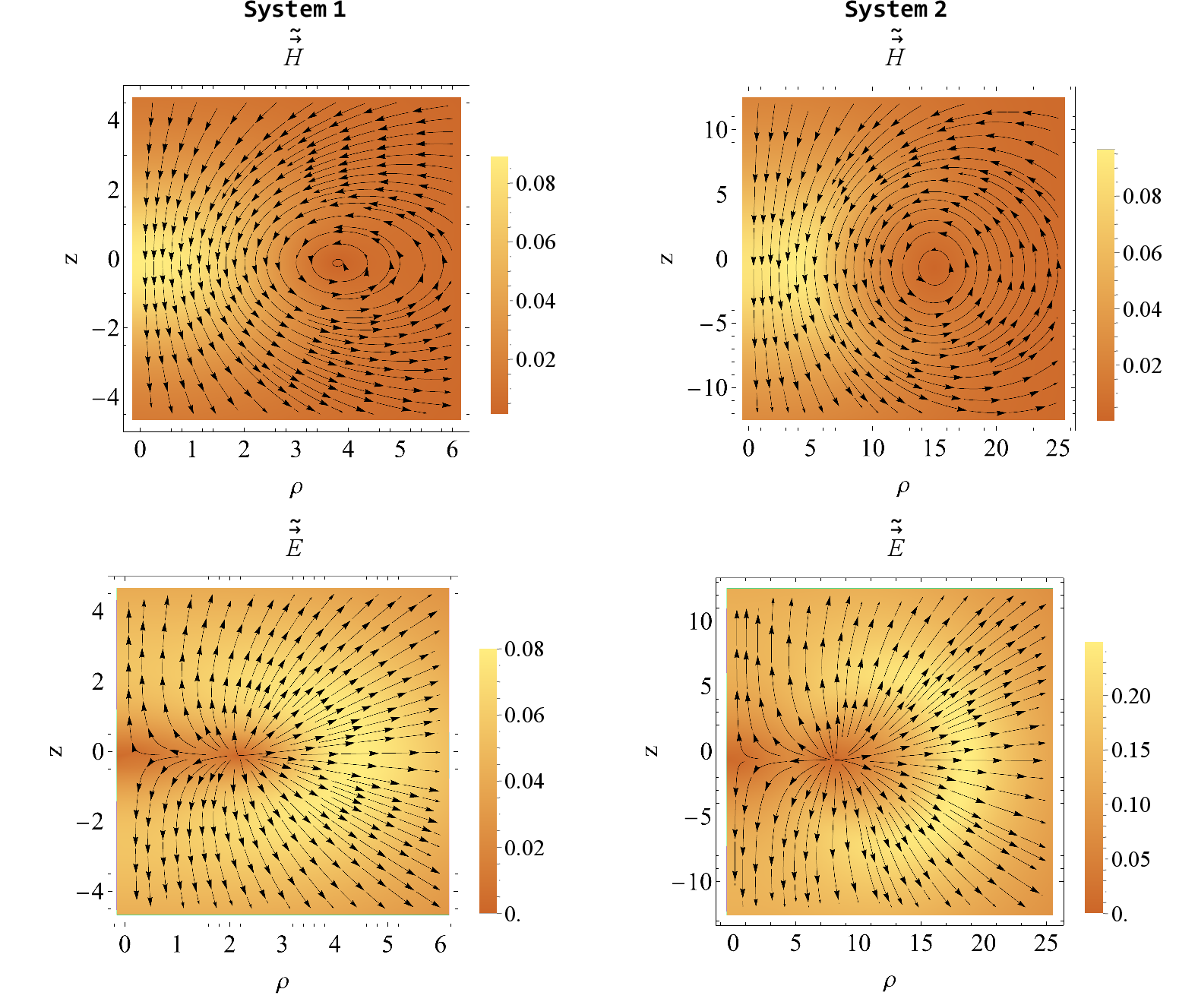}
     \end{center}
    \vspace{-1.cm}
    \caption{Dimensionless magnetic $\tilde{\vec{H}}\equiv \vec{H}/m_s^2$  and electric $\tilde{\vec{E}}\equiv \vec{E}/m_s^2$ field orientation of the configurations
    labeled by the dots 1 and 2 on the curves in Fig.~\ref{fig_M_freq_sf_sp}. The plots are made in a meridional plane $\varphi=\text{const.}$
spanned by the coordinates $\rho=x \sin\theta$ and $z=x \cos\theta$.
}
    \label{fig_EM_strength}
\end{figure}

The $U(1)$ gauged Q-balls have been earlier discussed in the literature (see, e.g., Refs.~\cite{Lee:1988ag,Lee:1991bn,Kusenko:1997vi,Gulamov:2015fya,Gulamov:2013cra,Panin:2016ooo,Nugaev:2019vru,Loiko:2019gwk}).
These solutions appear to exist up to a maximal value of the gauge coupling constant only because of an additional repulsive interaction arising from the gauge sector.  Typically, for a fixed value of the gauge coupling $e$, there are two branches of solutions. First, the lower in
energy branch emerges smoothly from the perturbative spectrum, as the frequency~$\tilde \omega$ is decreasing below the mass threshold.
This  branch bifurcates with the second, higher in energy branch of solutions
at some minimal value of the frequency $\tilde \omega_{\text{min}}$ below which no solutions are found. The second branch extends backwards as the frequency increases.

A single fermion state can be localized by the electromagnetic field of the Q-ball, however its contribution to the total energy is expected to be very small. In other words, the backreaction of the fermions on the Q-ball is almost negligible. Indeed, we have explored the solutions of the equations
\eqref{Dirac_eq_X}-\eqref{sf_eq_phi} by varying the field frequency\footnote{Clearly, in the context of $U(1)$ gauge theory, the frequency  $\tilde \omega_{\text{min}}$ is not an appropriate physical quantity but rather a useful parameter in our numerical simulations.} $\tilde \omega_{\text{min}}$ and found such configurations. Since these equations depend on five parameters,
$\tilde\omega, \tilde m, a, \tilde{b}$, and $e$,
we shall not pursue a complete scanning of the domain of existence of these solutions but rather  illustrate a general pattern setting the scaled bare fermion mass  $ \tilde m=1$ and the gauge coupling $e=0.1$  for simplicity. Hereafter the parameters of the scalar potential are fixed as  $a=3.8, \tilde{b}=3$.

In Fig.~\ref{fig_M_freq_sf_sp}, the scaled total mass $\tilde M$ from Eq.~\eqref{M_tot} and the total angular momentum of the solutions $\tilde J$ from Eq.~\eqref{ang_mom_tot} are exhibited versus the
scaled  frequency $\tilde\omega$. Similar to the
case of the gauged Q-balls without fermions, there are two branches of solutions which exist within the interval $\tilde\omega \in
[\tilde\omega_{\text{min}}; 1]$, and for $e=0.1$ the minimal value of the frequency is $\tilde\omega_{\text{min}} \approx 0.68$. Further, the energy, the charge, and the total angular momentum of the  configuration are minimal at some critical value of the frequency $\tilde\omega_{\text{cr}} \approx 0.92 > \tilde\omega_{\text{min}}$ (see Fig.~\ref{fig_M_freq_sf_sp}, the upper left panel). Hence, the Vakhitov-Kolokolov stability criterion \cite{Vakhitov:1973lcn} indicates that  classically stable solutions may exist for $\tilde\omega \in
[\tilde\omega_{\text{min}}; \tilde\omega_{\text{cr}}]$. Note that
the energy, the charge, and the total angular momentum
of the gauged Q-balls remain finite in the limit $\tilde\omega \to 1 $. Further, our numerical results confirm that the contribution of the spinor field to the total energy always remains extremely small compared to the contributions coming from the scalar and electromagnetic fields, respectively. Indeed,
the normalization condition \eqref{norm} implies that we consider one-fermion spin-1/2 state  localized by the electromagnetic field of a charged scalar condensate. However, there is a relatively small effect of the presence of localized fermions
in the system: we observe that  the minimal value $\tilde \omega_{\text{min}}$  is slightly decreasing as compared to the pattern of evolution of the pure gauged Q-ball.

In the upper right panel of Fig.~\ref{fig_M_freq_sf_sp}, we exhibit the dependence of the eigenvalues of the Dirac operator $\tilde{\Omega}$ on the frequency of the  scalar field  $\tilde \omega$. We observe that a localized fermionic mode exists for the entire range of allowed values of  $\tilde \omega$. On the lower branch this mode is very close  to the positive continuum, as seen in Fig.~\ref{fig_M_freq_sf_sp}. The spectral flow of the fermionic Hamiltonian bifurcates at $\tilde\omega_{\text{min}}$, the eigenvalue $\tilde{\Omega}$ is rapidly decreasing on the second branch. However, it always remains positive, and there is no zero mode in the spectrum.

Numerical solution of the equations \re{1_10}-\re{1_25} shows that the components $\psi_{1,2}$ of the spinor~\eqref{spinor} are  multivalued functions at the origin. In particular, $\psi_1\left.\right|_{x=0}=\left| \psi_1\right| e^{\imath \arg(\psi_1)}$, where
$\arg(\psi_1)=\arctan\left[(V+W)/(X+Y)\right]$ is a multivalued function of polar angle at the origin. One can show that the usual transformation of the components of the spinor~\eqref{spinor}  from the spherical to the Cartesian basis leads to the functions which are single-valued everywhere in space (see Appendix \ref{App}). Clearly, the fermionic density $\tilde{\psi}^\dag \tilde{\psi}$ is regular and single-valued for any choice of the basis.

The characteristic size of the configuration increases as the frequency $\tilde\omega$ is decreasing along the lower branch. Consequently, both the scalar current \re{1_35} and associated magnetic field become stronger.
As the bifurcation
with the second, higher-energy branch is approached, the contribution of the electrostatic repulsion to the total energy affects the fine energy balance between the volume and surface energies of the scalar condensate.
When $\tilde\omega$ is increased again along the second branch, the volume of the configuration continues to
increase (cf. Figs.~\ref{fig_fields_distr} and~\ref{fig_fields_distr_equator} where the corresponding field distributions are  displayed).

Let us now consider the contributions of the electric and magnetic fields to the total energy density of the configuration \eqref{T_tt}. It can be evaluated as
\begin{equation}
    \left(\tilde{T}_t^t\right)_{\text{EM}} \equiv \left(\tilde{T}_t^t\right)_{\text{E}}+\left(\tilde{T}_t^t\right)_{\text{M}}=
 \frac{1}{2}\left(
        \tilde \chi_{,x}^2 + \frac{1}{x^2}\tilde \chi_{,\theta}^2 \right)
+\frac{1}{2} \left(\frac{\csc^2\theta }{x^2} \sigma_{,x}^2
                + \frac{\csc^2\theta }{x^4} \sigma_{,\theta}^2
            \right)  . \nonumber
\label{T_tt_EM}
\end{equation}

Thus,  the scaled electric  field energy is
$$
\tilde W_{\text{E}}\equiv W_{\text{E}}/m_s=2\pi \int_{0}^{\infty}dx\int_{0}^{\pi}d\theta \left(\tilde{T}_t^t\right)_{\text{E}} x^2 \sin\theta  ,
$$
and the scaled magnetic field energy is
$$
\tilde W_{\text{M}}\equiv W_{\text{M}}/m_s=2\pi \int_{0}^{\infty}dx\int_{0}^{\pi}d\theta \left(\tilde{T}_t^t\right)_{\text{M}} x^2 \sin\theta  .
$$

To illustrate these quantities, we display in the lower row of Fig.~\ref{fig_M_freq_sf_sp} the  total electromagnetic energy and the ratio of the energies of  electric and magnetic fields versus the frequency~$\tilde \omega$. As we see from the left panel of this figure, the   contribution of the energy of electromagnetic field to the total energy of the system becomes more significant along the  upper branch, whereas, on the lower branch, this contribution gradually decreases as the frequency $\tilde \omega$ is increasing up to the  critical value $\tilde \omega_{\text{cr}}$. It slowly increases again as the frequency increases further up to the mass threshold. In turn, the contribution of the electric energy is always higher than the corresponding energy of the magnetic field (see
 Fig.~\ref{fig_M_freq_sf_sp}, the lower right plot). Thus,
 on both branches  the dominating contribution is provided by the electric field
that ensures the electrostatic repulsion required for the equilibrium of the system.

In Fig.~\ref{fig_fields_distr}, we display the 3D plots of the scalar field $\tilde \Phi$, electric potential $\tilde \chi$, magnetic potential $\sigma$, and fermionic density $\tilde \psi^\dag \tilde\psi$
for the corresponding solutions on two different branches for the same value of the scalar frequency $\tilde\omega=0.82$; the corresponding configurations are labeled by the dots on the curves in Fig.~\ref{fig_M_freq_sf_sp}. Further,
in Fig.~\ref{fig_EMT_currents}, we exhibit the corresponding distributions of the total energy density from Eq.~\eqref{T_tt}, the physical component of the angular momentum density from Eq.~\eqref{T_tphi}, the total charge density $\tilde{j}^t= \tilde{j}^t_\psi+\tilde{j}^t_\phi$, and the physical component of the total current density
$\tilde{j}^\varphi=\tilde{j}^\varphi_\psi+\tilde{j}^\varphi_\phi$ evaluated from Eqs.~\eqref{1_30} and~\eqref{1_35} as  functions of the cylindrical coordinates $\rho=x \sin\theta$ and $z=x \cos\theta$.

Our simulations show that along the lower branch a single peak appears in the corresponding distributions (see  Fig.~\ref{fig_EMT_currents}, the left column). Also, the magnitude of the scalar field~$\tilde \Phi$ possesses a single maximum on the symmetry axis (see Fig.~\ref{fig_fields_distr}, the upper left plot). The maximal value of $\tilde \Phi$ on this branch usually remains below the corresponding false vacuum of the potential~\re{U_sf}, except the configurations located near the turning point
(see the upper left plot of Fig.~\ref{fig_fields_distr_equator}).

When we inspect the fields along the upper  branch, we note that the Q-ball is transformed into the thin-wall regime: instead of a single maximum,
a plateau of an almost constant value of $\tilde \Phi$, which is slightly above the false vacuum,
is formed in some domain (see the upper right plot of Fig.~\ref{fig_fields_distr} and the upper left plot of Fig.~\ref{fig_fields_distr_equator}).
The corresponding distribution of the total energy density is volcano-shaped,  with a domain wall that is separating the vacuum on the exterior and local minimum in the interior
(see Fig.~\ref{fig_EMT_currents}, the right column).

The profiles of  typical solutions in the equatorial plane are shown in
Fig.~\ref{fig_fields_distr_equator}. Here we display the radial dependencies of the functions $\tilde \Phi$, the magnitudes of the electric  ($|\tilde E|$) and magnetic ($|\tilde H|$) fields,
the density of the fermionic state, and the densities of the total energy and angular momentum. Clearly, in the thin-wall regime, the fields are almost confined in the interior of the Q-ball with the long-range Coulomb electric field outside.

The $j^\varphi$ components of the currents~\re{1_30} and~\re{1_35} generate a  toroidal magnetic field $\vec{H}$, which forms a vortex encircling the configuration (see Fig.~\ref{fig_EM_strength}). The size of the vortex rapidly increases along the second branch, and it can be effectively decomposed into two
magnetic fluxes, one of which encircles the Q-ball in equatorial plane, and the second one is directed along the
symmetry axis.

The electric field of the configuration $\vec{E}$ corresponds to the field of a toroidal distribution of a positive electric charge with the center on the circle
$\{\rho=\rho_0, z=0\}$ and with a single node at the origin (see  Fig.~\ref{fig_EM_strength}). The  characteristic size of this domain is increasing along the second branch,
although the qualitative nodal structure of the electric field remains the same. From these plots, as well as from the graphs for
$|\tilde E|$ and $|\tilde H|$ shown in Fig.~\ref{fig_fields_distr_equator}, it is also seen that the maximal value of the magnetic field increases along the first branch (i.e., as the frequency $\tilde{\omega}$ decreases).
However, along the second branch, as  $\tilde{\omega}$ is increasing,  the maximal value of the magnitude of the magnetic field decreases again. For the magnitude of the electric field, the situation is different: it always increases both along the first and the  second branches.

In both branches, the energy of the magnetic field always remains smaller than the electrostatic energy (see the lower right panel of Fig.~\ref{fig_M_freq_sf_sp}). For this reason,
the fermionic mode can be thought of as being localized by the electric field of a Q-ball.
The distribution of the fermionic density of the localized mode on the first branch is almost spherically symmetric (see Fig.~\ref{fig_fields_distr}, the lower left plot). The spacial distribution of the scalar condensate, however, is of a toroidal shape.

\section{Conclusion}
\label{conclus}

The main result in this paper is that
charged Dirac fermions can be localized by the electromagnetic field of  $U(1)$ gauged Q-balls.  
The full system of field equations for the complex charged scalar field and the
Dirac fermions indirectly coupled to the Q-ball via the minimal electromagnetic
interaction is supplemented by the normalization condition for the localized fermions.
Imposing appropriate boundary conditions, we constructed numerical solutions of the resulting system of
integral-differential equations and found the corresponding energy eigenvalues.
We have shown that the bounded single mode exists for the entire allowed range
of parameters of the model,  although the backreaction of this mode is very small.

By analogy
with the case of pure $U(1)$ gauged Q-balls, there are two branches of solutions which bifurcate at some minimal value of the scalar frequency, which increases as the gauge coupling increases. On the other hand, the presence of the localized fermions
has the opposite effect: for a given value of the gauge coupling the  minimal value of scalar frequency is slightly decreasing.

The work here should be taken further by considering other normalizible fermionic states localized by the gauged Q-ball, and on the upper branch they would correspond to higher Landau levels. Another interesting
question which we hope to address in the near future is to investigate the Dirac fermions localized on the Q-ball via the Yukawa coupling.

\section*{Acknowledgements}

V.D. and V.F. gratefully acknowledge support provided by the program
No.~BR24992891 (Integrated research in nuclear, radiation physics and engineering, high energy physics and cosmology for the development of competitive technologies)
of the Ministry of Education and Science of the Republic of Kazakhstan. Y.S. gratefully acknowledges the support by FAPESP,
project No 2024/01704-6 and thanks the Instituto de F\'{i}sica de S\~{a}o Carlos; IFSC for kind hospitality.

\appendix

\section{Operator of the total angular momentum in spherical coordinates}
\label{App}
It is well known that a transformation of Dirac spinors from  Cartesian to spherical  coordinates, or vice versa, is defined as
\begin{equation}
    \psi = \mathcal{G} \psi^\prime , \nonumber
\label{app_10}
\end{equation}
where $\mathcal{G}$ is a transformation matrix. The corresponding spin operator $\vec S$ transforms as
\begin{equation}
    S_i =  \mathcal{G} S^\prime_i  \mathcal{G}^\dagger  . \nonumber
\label{app_20}
\end{equation}

Explicitly, the matrix $\mathcal{G}$ which transforms spinors \eqref{spinor} from Cartesian to spherical coordinates has the form
\begin{equation}
    \mathcal{G} = \frac{1}{\sqrt{2}}
    \begin{pmatrix}
        e^{\frac{\imath}{2} (\theta +\varphi )} & -\imath e^{\frac{\imath}{2} (\theta -\varphi )} & 0 & 0 \\
        e^{-\frac{\imath}{2} (\theta -\varphi )} & \imath e^{-\frac{\imath}{2} (\theta +\varphi )} & 0 & 0 \\
        0 & 0 & e^{\frac{\imath}{2} (\theta +\varphi )} & -\imath e^{\frac{\imath}{2} (\theta -\varphi )} \\
        0 & 0 & e^{-\frac{\imath}{2} (\theta -\varphi )} & \imath e^{-\frac{\imath}{2} (\theta +\varphi )}
    \end{pmatrix}   . \nonumber
\label{app_30}
\end{equation}
The corresponding operator of the  angular momentum in spherical coordinates $\hat{M_i}$  becomes
\begin{equation}
    \hat{M}_i = S_i - \imath D_i , \nonumber
\label{app_40}
\end{equation}
where $D_i$ is a covariant derivative of the spherical spinor.
In particular, the $S_\varphi$ component of the spin operator in spherical coordinates has the form
\begin{equation}
    S_\varphi = \frac{1}{2}
    \begin{pmatrix}
        0 & e^{\imath \theta } & 0 & 0 \\
        e^{-\imath \theta } & 0 & 0 & 0 \\
        0 & 0 & 0 & e^{\imath \theta } \\
        0 & 0 & e^{-\imath \theta } & 0
    \end{pmatrix}   . \nonumber
\label{app_50}
\end{equation}

\end{document}